\documentclass[apjl]{emulateapj}
\usepackage{graphicx, graphics}
\usepackage{amsmath,amssymb}

\received{}
\revised{}
\accepted{}

\newcommand{\chandra}{\textsl{Chandra}}
\newcommand{\xmm}{\textsl{XMM}-Newton}
\newcommand{\kms}{$\rm km\ s^{-1}$}
\newcommand{\xspec}{{\sc xspec}}
\newcommand{\ovii}{\rm O~\uppercase\expandafter{\romannumeral7}}

\shorttitle{X-Ray Absorption by \ion{O}{7} in the Hercules Supercluster}
\shortauthors{Ren, Fang, \& Buote}

\begin{document}
\title{X-Ray Absorption by the Warm-Hot Intergalactic Medium in the Hercules Supercluster}

\author{Bin~Ren\altaffilmark{1,2}, Taotao~Fang\altaffilmark{1,3}, David~A.~Buote\altaffilmark{3}}

\altaffiltext{1}{Department of Astronomy and Institute of Theoretical Physics and Astrophysics, Xiamen University, Xiamen, Fujian 361005, China; fangt@xmu.edu.cn}
\altaffiltext{2}{Department of Physics, Xiamen University, Xiamen, Fujian 361005, China}
\altaffiltext{3}{Department of Physics \& Astronomy, 4129 Frederick Reines Hall, University of California, Irvine, CA 92697, USA}

\begin{abstract}

The ``missing baryons'', in the form of warm-hot intergalactic medium (WHIM), are expected to reside in cosmic filamentary structures that can be traced by signposts such as large-scale galaxy superstructures. The clear detection of an X-ray absorption line in the Sculptor Wall demonstrated the success of using galaxy superstructures as a signpost to search for the WHIM. Here we present an {\sl XMM}-Newton Reflection Grating Spectrometer (RGS) observation of the blazar Mkn~501, located in the Hercules Supercluster. We detected an \ion{O}{7} K$\alpha$ absorption line at the 98.7\% level ($2.5\sigma$) at the redshift of the foreground Hercules Supercluster. The derived properties of the absorber are consistent with theoretical expectations of the WHIM. We discuss the implication of our detection for the search for the ``missing baryons''. While this detection shows again that using signposts is a very effective strategy to search for the WHIM, follow-up observations are crucial both to strengthen the statistical significance of the detection and to rule out other interpretations. A local, $z\sim0$ \ion{O}{7}  K$\alpha$ absorption line was also clearly detected at the $4\sigma$ level, and we discuss its implications for our understanding of the hot gas content of our Galaxy.

\end{abstract}

\keywords{BL Lacertae objects: individual (Markarian 501) --- cosmology: observations --- diffuse radiation --- X-rays: diffuse background --- X-rays: galaxies: clusters} 

\section{Introduction}
There is an apparent baryon deficit predicted by the big bang nucleosynthesis and the cosmic microwave background radiation observations (e.g., \citealp{bristow1994, fukugita1998}). Latest numerical simulations and observations suggested that between 30 -- 50\% baryons are missing (see, e.g., \citealp{cen2006, shull2012}), and a significant amount of this ``missing baryons" are located in between galaxies, in the form of the so-called ``warm-hot intergalactic medium", or WHIM.

Due to its high temperature, the WHIM gas is mainly detectable in the ultraviolet (UV) and X-ray. While the first evidence of the gas is from the UV absorption lines in the spectra of background galactic active nuclei (AGNs) (see, e.g., \citealp{savage1998, tripp2000, oegerle2000, sembach2004, richter2004, danforth2005, stocke2006, lehner2007, howk2009}), the majority is expected to be detectable only in X-rays (e.g., \citealp{cen1999, dave2001, cen2006, shull2012}), which currently can be obtained with high resolution X-ray spectrometers on-board \chandra\ and \xmm.

In the past decade, several detections of highly ionized metals in the WHIM in X-ray have been reported (see, e.g., \citealp{fang2002, mathur2003, mckernan2004, fujimoto2004, nicastro2005}). However, most of these detections either have low statistical significance or cannot be independently confirmed by different instruments (see, e.g., \citealp{rasmussen2006, kaastra2006, fang2007, williams2006, yao2009}). A drawback of previous studies is that the background targets were selected based on their X-ray flux, i.e, the location of the WHIM is not known {\it a priori}. The detection significance of the any potential absorption would be lower since one has to search the entire sightline to rule out random fluctuations \citep{kaastra2006}. 

An improved technique, searching for the WHIM signature from known signposts, was proposed to improve the statistical significance. One of the the best candidates for such signpost is the large-scale galaxy superstructure, as predicted by cosmological simulations \citep{springel2003}. With \chandra\ and \xmm, \citet{buote2009} implemented this strategy and reported the detection of an \ion{O}{7} absorber resided in the Sculptor Wall Superstructure along the sightline toward the background blazar H~2356-309. This detection was later confirmed with improved statistics by \citet{fang2010} with a longer {\sl Chandra} observation. \citet{nicastro2013} also reported the detection of highly ionized metal absorption line systems in the \chandra\ spectrum of the blazar 1ES~1553+113 that have associated UV absorption lines. These detections highlight the success and effectiveness of using foreground signposts as tracers for the WHIM.

In this work, we extend our previous study of H~2356-309 and perform an X-ray observation of another blazar, Mkn 501, which is located in the Hercules Supercluster. Our main goal is to assess whether our detection of the WHIM in the Sculptor Superstructure is typical. The X-ray flux of Mkn~501 is slightly higher than those of H~2356-309. If the galaxy superstructures host significant amount of the WHIM as we detected in the Sculptor superstructure, we expect to detect similar feature in the Hercules Supercluster.

Our second motivation is to study the local hot gas around the Milky Way, either in the Galactic disk or in the distant halo in the form of the circum-galactic medium (CGM). Highly ionized metals in such hot gas will also leave its imprint in the spectrum of the background AGNs. A number of highly ionized, $z\sim0$ metal absorption systems (primarily \ion{O}{7}) have been reported along the sightlines toward background AGNs (see, e.g., \citealp{ nicastro2002, fang2003, rasmussen2003, mckernan2004, williams2005, bregman2007, gupta2012}). Such metal absorption systems may have a larger covering fraction (\citealp{fang2006, bregman2007, gupta2012, miller2013}), and we should be able to detect them regardless the direction, as long as the background source is bright enough \citep{fang2006}. This observation is an excellent opportunity to test this idea.

Our paper is arranged as follows. In section \S2 we describe our observation and data analysis. We discuss the implication of both the redshifted and the Milky Way \ion{O}{7} K$\alpha$ lines in Section \S3 and systematic errors in section \S4. We summarize our results in the last section.

\begin{figure}[t]
\center
\includegraphics[width=0.3\textwidth,height=.195\textheight,angle=0]{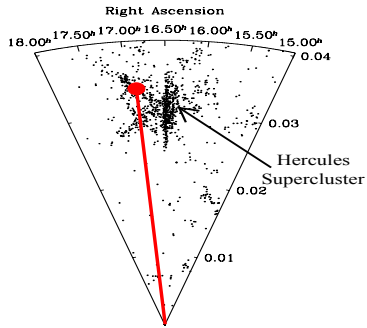}
\caption{Wedge diagram of the Hercules Supercluster and Mkn 501 sightline (red line), in the RA direction. This diagram covers the Dec direction with $37^{\circ}<\delta<42^{\circ}$. The red dots are the location of Mkn 501. Data were taken from the NASA/IPAC Extragalactic Database.}
\label{f1}
\end{figure}

\begin{deluxetable}{ccccc}
\tablecaption{\xmm\ Observation Log\label{t:log}}
\tablewidth{0pt}
\tablehead{
\colhead{Observation ID}& \colhead{Exposure} & \colhead{Date} & \colhead{$\Gamma^{a}$} & \colhead{Flux$^{b}$}\\
\colhead{ }&\colhead{(ks) }&\colhead{}&\colhead{} &\colhead{}}
\startdata
0652570101 & 44.9 & 2010 Sep 09 & $2.45\pm0.01$ & 2.92 \\
0652570201 & 44.9 & 2010 Sep 11 & $2.44\pm0.01$ & 3.06 \\
0652570301 & 40.9 & 2011 Feb 11 & $2.17\pm0.01$ & 3.21 \\
0652570401 & 40.7 & 2011 Feb 15 & $2.08\pm0.01$ & 4.27
\enddata
\tablecomments{a. Unless specified, errors are $1\sigma$ in this paper. b. Flux between 0.5 and 2 keV, in units of $10^{-11}\rm\ erg\ s^{-1}\ cm^{-2}$.}
\end{deluxetable}

\section{Observation \& Data Analysis}
\label{datareduction}

Hercules Supercluster is a superstructure that extends for more than 100 Mpc from $z\sim0.03$ to 0.04 \citep{abell1961, einasto2001, kopylova2013}. This supercluster contains a few ten large, Abell galaxy clusters. Figure~\ref{f1} shows the wedge diagram of the Hercules Supercluster between RA=$15^h$ and $18^h$, in which Mkn 501 (the red dot at $z=0.0337\pm0.0001$, \citealp{devaucouleurs1991}) is embedded. Because of its strong X-ray flux, Mkn~501 provides an excellent opportunity to study the WHIM gas that may be located in the foreground superstructure. We performed four observations of Mkn 501, taken with the Reflection Grating Spectrometer (RGS) on \xmm\ in 2010 September and 2011 February. The total exposure time is $\sim 170$ $ksec$ (see Table~\ref{t:log} for details). Although there are two identical sets of gratings in RGS (i.e., the RGS1 and RGS2), in this paper we only focus on RGS1, since RGS2 is not sensitive over the energy range of \ion{O}{7} K$\alpha$ line due to its CCD failure. 

We generated the source spectra, the background files, as well as the response matrices for the RGS1 data with \xmm\ Science Analysis Software (SAS Version 13.0.0)\footnote{http://xmm.esa.int/sas/}, using the latest calibration files. We also checked the background events by generating a light curve for CCD9, which is most susceptible to proton events, and found no significant flares. We restricted our analysis to the first order spectra. Finally, to increase photon statistics and optimize the detection of absorption lines, we rebinned the spectra to have a minimum of 30 counts per bin.

\begin{figure*}[t]
\center
\includegraphics[width=0.49\textwidth, height=0.195\textheight]{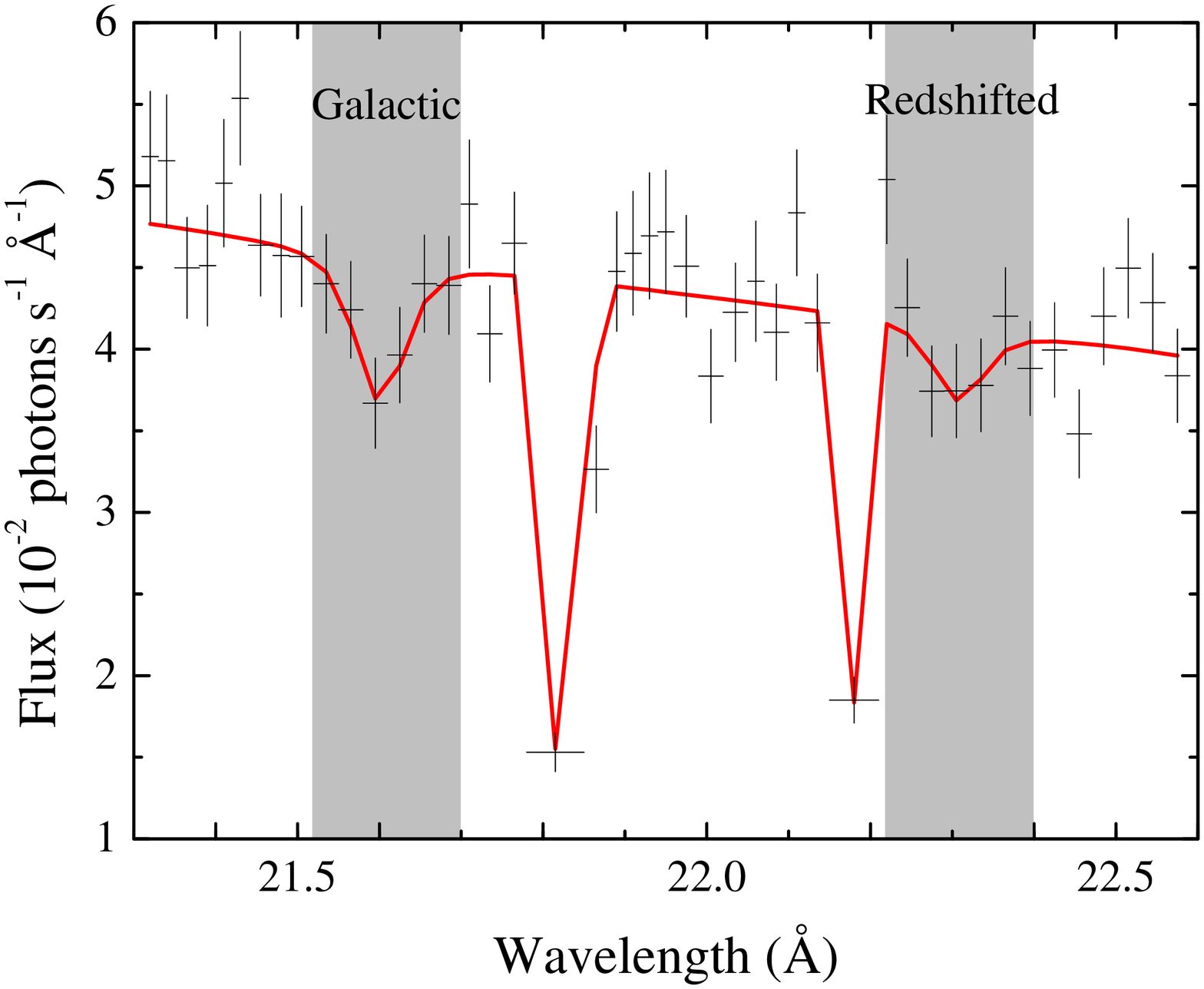}
\includegraphics[width=0.49\textwidth, height=0.195\textheight]{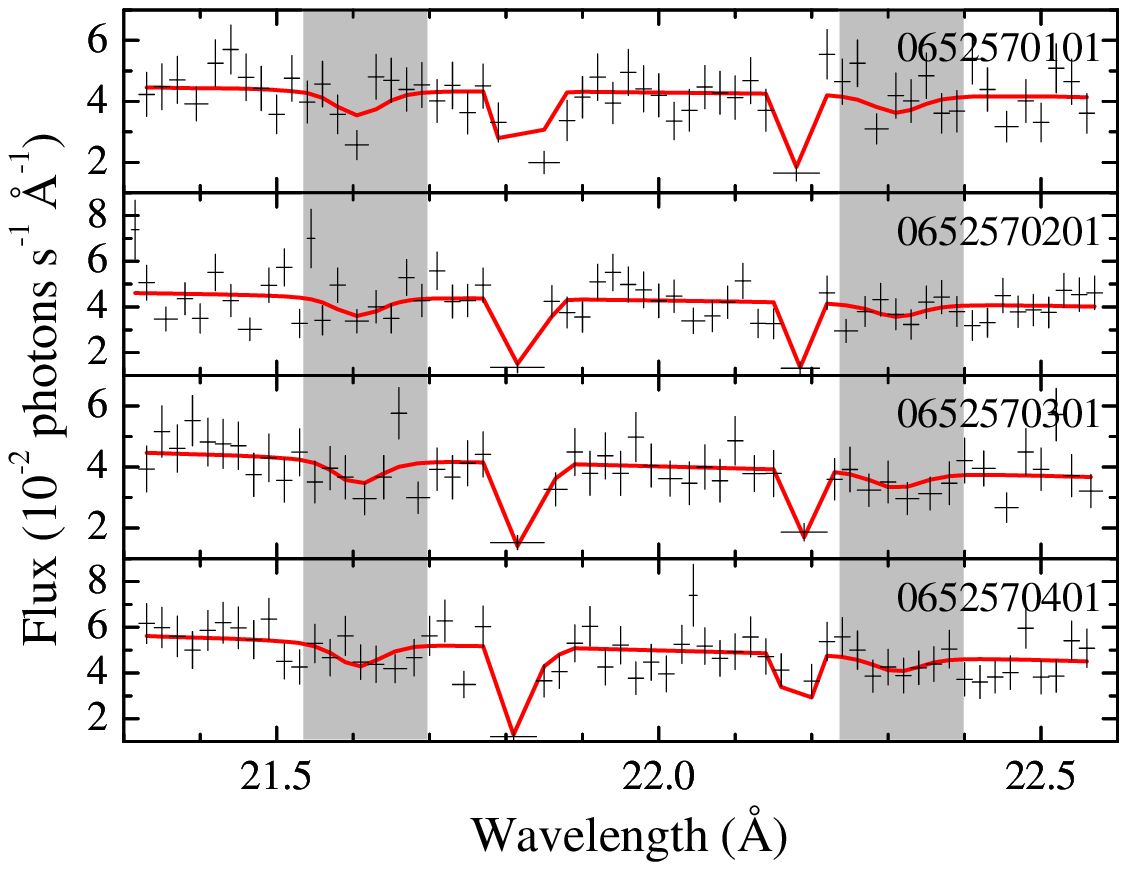}
\caption{Left panel: Stacked spectrum ($21.3$--$22.6$ \AA) of the four observations. The red line is the model spectrum, the two dips around $21.6$\AA\ and $22.3$ \AA\ are absorption features focused in this paper. Features at $\sim21.80$ \AA\ and $\sim22.20$ \AA\ are instrumental. Note: This spectrum is only for demonstration purpose. Right panel: \xmm\ datasets, labeled with their observational IDs used in spectral fitting.}
\label{f2}
\end{figure*}

The processed data were analyzed with \xspec (version 12.8.1)\footnote{http://heasarc.gsfc.nasa.gov/xanadu/xspec/}. We adopt the value of $1.74\times10^{20}$ cm$^{-2}$ for the column density of the foreground absorbing neutral Hydrogen in the Milky Way \citep{dickey1990}. We also adopt the Cash C-statistic \citep{cash1979} for an unbiased estimate \citep{humphrey2009} for the model parameters. We find the broadband continuum can be fitted very well with a power law model plus the Galactic neutral hydrogen absorption. In Table~1 we list the observational log and the best-fit continuum parameters: The power law photon index ($\Gamma$) and the flux between 0.5 and 2 keV. The flux varied by $\sim$ 30 --- 50\% during a time span of six months between September 2010 and February 2011.

We then restricted our fit in a narrow band between 21.3 and 22.6 \AA, in which the local $z=0$ and the Hercules \ion{O}{7} absorption are expected to present. We fitted the absorption feature with a spectral model based on the Voigt line profile (see Buote et al.~2009 and Fang et al.~2010 for model details). Figure~\ref{f2} left panel shows the stacked spectrum. We emphasize that the stacked spectrum is only for display purpose. The reason we did not fit the stacked spectrum is because such procedure is likely to introduce artificially narrow spectral features due to the difference in the response matrix of each observation (see \citealp{rasmussen2006} for details). The actual fitting was performed simultaneously on four observations, and we tied the physical properties, the column density, redshift, and the Doppler-$b$ parameter, together among the observations. The data and model of each observation are plotted in the right panel of Figure~2.  We clearly detected the local $z=0$ \ion{O}{7} absorption line at 21.6 \AA. The equivalent width (EW) of this line is $22.3\pm5.7$ m\AA,\footnote{Errors are $1\sigma$ in this paper unless otherwise specified.} indicating a line significance of $\sim 4\sigma$. We estimated the error on EW using Monte-Carlo simulation (see \citealp{buote2009} and \citealp{fang2010} for details). We examined the shorter wavelength but did not find a significant presence of the corresponded \ion{O}{7} K$\beta$ and K$\gamma$ lines.

\addtolength{\tabcolsep}{-1.5pt}
\begin{deluxetable}{ccccc}
\tablecaption{\ion{O}{7} Absorption Line Properties\label{t:line}}
\tablewidth{0pt}
\tablehead{
\colhead{Line}& \colhead{log($N^a$)} & \colhead{Doppler-$b$} & \colhead{Redshift} & \colhead{EW}\\
\colhead{ }&\colhead{} & \colhead{($\rm km\ s^{-1}$)} & \colhead{} &\colhead{(m\AA)}}
\startdata
Milky Way & $16.68_{-0.81}^{+1.50}$ & $80^{+307}_{-60}$ & $0.0002\pm0.0004$ & $22.3\pm5.7$ \\
Redshifted & $15.86_{-0.24}^{+2.37}$ & ... & $0.0329\pm0.0008$ & $18.8\pm6.5$
\enddata
\tablecomments{a. Column density in units of $\rm cm^{-2}$.}
\end{deluxetable}

To search for the WHIM feature from the Hercules Supercluster, we examined the spectra between 22.25 and 22.35 \AA\ where the redshifted \ion{O}{7} K$\alpha$ line may be located. Indeed, we find a weak absorption feature at $\sim22.3$ \AA.\ By including an absorption line there, the $C$-statistic improves by $\Delta C = 7.2$. This feature, if identified as an \ion{O}{7} K$\alpha$ line, would have a redshift of $z=0.0329\pm0.0008$, right within the Hercules superstructure. We fitted the line with the same Voigt-profile based model. The best-fit line parameters are listed in Table~\ref{t:line}. We cannot constrain the Doppler-$b$ parameter, so we let it vary between 20 and 600 $\rm km\ s^{-1}$. The low limit is adopted by assuming this line is purely thermally broadened, and the upper limit by assuming shock-wave broadening in the IGM \citep{kang2005}. We followed the procedure introduced in \citet{buote2009} and employed a Monte Carlo procedure using C-statistic to estimate the detection significance. We first adopted the best-fit parameters for the four continua (i.e., spectra without the absorption feature), and generated $10,000$ sets of faked spectra. We then performed the fit using the exact same procedure that we applied for the observed data. We estimated the change of the $C$-statistics, $\Delta C$, between the model fit with and without the additional absorption line. Based on our {\it a priori} knowledge of the expected feature, we restricted our search for the absorption line in the redshift range of 0.03 --- 0.0337. We then compared the $\Delta C$ in the faked data with the real one. Among the $10,000$ simulations, we found 132 false ones with a change equal or larger than the observed $\Delta C$. This suggests a detection significance of 98.7\%, or 2.5$\sigma$.

\section{Discussion of the \ion{O}{7} Absorbers}

\subsection{The Redshifted Absorber}

Our detection of the Hercules X-ray absorber allows us to make a rough estimate of temperature and density of the absorber, assuming it is produced by the WHIM residing in the superstructure. Since we do not detect absorptions from other ion species at similar redshift, the temperature is likely to be around $10^6$ K for maximum \ion{O}{7} ionization fraction, assuming collisional ionization. The best-fit \ion{O}{7} column density is $7.5\times10^{15}$~cm$^{-2}$. Following \citet{fang2010} and assuming that the absorber is uniformly distributed between $z=0.03$ and 0.034 with a path length of $\sim 15$ Mpc, the density of hydrogen is
{
\small
\begin{eqnarray*}\label{eq:1}
n & = & 4 \times10^{-6}\ (\rm cm^{-3}) \nonumber \\
  &   & \times \frac{N_{\ovii}}{7.5\times10^{15}\rm\ cm^{-2}}
    \left(\frac{f_{\ovii}}{1}\right)^{-1}
    \left(\frac{Z}{0.1Z_{\odot}}\right)^{-1} \left(\frac{L}{15\rm\
    Mpc}\right)^{-1}\hspace{-5pt},
\end{eqnarray*}
}where $N_{\ovii}$ is the column density of \ion{O}{7}, $f_{\ovii}$ is the \ion{O}{7} ionization fraction, $L$ is the length of the absorber, and we assume a metallicity, $Z$, of $0.1$ solar abundance, as suggested by cosmological simulations (e.g., \citealp{cen1999, shull2012}). Therefore, the number density of hydrogen corresponds to an overdensity of $\delta \approx 20$. Here $\delta = n/n_0 -1$ where $n_0$ is the mean density of the universe. This value is similar to what we detected in the Sculptor Wall absorber (\citealp{buote2009, fang2010}) and is fully consistent with the predictions from cosmological simulations ($\delta=5$--$200$; see, e.g., \citealp{dave2001}).

The Hercules supercluster has a complex environment along the Mkn~501 sightline, including galaxies, galaxy groups, and galaxy clusters. So it is also likely this absorber is located in one of these foreground systems. We have used the NASA/IPAC Extragalactic Database and the seventh data release of the Sloan Digital Sky Survey
(SDSS DR7, \citealp{abazajian2009}) to search for foreground systems along the Mkn~501 sightline. We define a search radius of 5', which corresponds to $\sim$ 200 kpc (typical virial radius for a Milky Way-size halo) at the redshift of the Hercules absorber.  We found six galaxies, one galaxy group and one galaxy cluster with redshifts between 0.029 and 0.034. 

It is unlikely the absorber is located in either the intracluster medium or the intragroup medium. The hot intracluster medium typically has a temperature of more than $10^7$ K, and oxygen would be completely ionized at this temperature. The galaxy group, located at $z=0.0332$, has a velocity dispersion of $\sigma=243\rm\ km\ s^{-1}$ \citep{berlind2006}. Using the $\sigma-T$ relation of \citet{xue2000}, the implied temperature of the intragroup medium is $\sim 6\times10^6$ K. At this temperature most \ion{O}{7} will be ionized to \ion{O}{8}.

We also examined the six galaxies within $\sim$ 200 kpc of the sightline. The $z=0$ absorber detected here may be located in either the halo or the disk of our Galaxy, so it is also likely a foreground galaxy produces similar absorption. We estimated their virial radius to access whether Mkn~501 sightline passes through the halo of these galaxies. Four of six foreground galaxies have been identified in the SDSS DR7 with $r$-band magnitude ranging from $\sim$ 15 to 18.56. Using the $M_r$-halo mass relation from \citet{tinker2009} we estimated minimum halo mass of these galaxies to be in the range $10^{11}$ to $5\times10^{12}\rm\ M_{\odot}$, corresponding to virial radius of 150 -- 450 kpc. So it is likely the Hercules absorber actually is located in one of these foreground galaxies. Longer exposure in X-ray, as well as spectroscopical observations in other wavebands, are necessary to constrain the physical properties of the absorbers and distinguish between a galaxy or an IGM origin of the X-ray absorber. 

In addition to a WHIM origin, it is possible that the redshifted absorption is intrinsic to Mkn 501. The redshift of the absorption line is $z=0.0329\pm0.0008$, while the redshift of the blazar is $z=0.0337\pm0.0001$. Therefore, the difference in redshift could be explained by an outflowing \ion{O}{7} with a speed of 0 --- 500 \kms. However, we consider it is likely not the case. Typically, X-ray absorption lines detected in AGNs are produced by metals in the outflows with multiple velocity components at various ionization stages (see, e.g., \citealp{kaspi2003}). We examined the spectrum and did not find other absorption lines at similar redshift. We also did not find significant variation of the line during the four observations that spanned a period of 6 months, i.e., the significance of the line  does not arise from a single observation. 

To further investigate the likelihood of an outflow, we examined the ultraviolet spectrum of this blazar. The Far Ultraviolet Spectroscopic Explorer ({\sl FUSE}) observation only revealed a few absorption lines local to the Milky Way \citep{wakker2003}. We also examined the data taken by the Goddard High Resolution Spectrograph (GHRS) on-board the Hubble Space Telescope (Charles Danforth, private communication). We only found a few intergalactic \ion{H}{1} Ly$\alpha$ absorption lines. For these reasons we believe that the redshifted \ion{O}{7} absorber is likely associated with the intervening gas in the foreground supercluster. Further observations with instrument that has high sensitivity  (such as the Cosmic Origins Spectrograph) may be necessary to detect outflow.  

We emphasize that it is unlikely the interstellar medium (ISM) in the host galaxy is responsible for this absorber. The stellar velocity dispersion of the host galaxy is $\sigma=372\pm18\rm\ km\ s^{-1}$ \citep{barth2002}. The implied virial temperature is $T_{vir} = \mu m_p\sigma^2/k \sim 1.4$ keV, where $\mu$ is the molecular weight, and $k$ is the Boltzmann constant. This temperature again is too hot for the production of \ion{O}{7}.

\subsection{The Milky Way Absorber}

As we expected, we clearly detected a zero-redhift \ion{O}{7} K$\alpha$ absorption line at $\sim$ 21.6 \AA.\ The line center is in a velocity range of -240 to 300 $\rm km\ s^{-1}$, suggesting this absorber is local to the Milky Way. The derived line EW, $22.3\pm5.7$ m\AA\, is comparable to those measured along other sightlines toward background AGNs (see, e.g., \citealp{bregman2007, gupta2012} and Fang et al., in preparation). The derived \ion{O}{7} column density of $\sim 5\times10^{16}\rm\ cm^{-2}$ suggests a large amount of hot ($T\sim10^6$ K), metal-enriched gas along this sightline in our Galaxy. In \citet{fang2006}, We presented a survey of \ion{O}{7} K$\alpha$ line in 20 AGNs, and found that this line was detected in all the high quality spectra. We suggested that the \ion{O}{7}-enriched hot gas likely to have a large sky covering fraction. The detection of such feature along the Mkn~501 sighltline confirms this idea.

The exact location of this hot gas is still unclear: It could be in the Milky Way disk in the form of the interstellar medium, or in the distant halo in the form of the so-called ``circum-galactic medium" (see, e.g, \citealp{fang2006, yao2007, gupta2012, miller2013}). The answer to this question will have important implication to our understanding of the baryon content of our Galaxy. Further surveys of AGN sightlines, along with X-ray emission observations and absorption line studies along the sightlines toward Galactic backgrounds such as the X-ray binaries will help resolve this problem.

\section{Systematic Errors}

The errors associated with our line measurement are largely dominated by the statistical errors, so we will not provide a detailed analysis of all the systematic errors. However, we briefly discuss several important ones that may have an impact on our measurement.

Our choice of bin size is to ensure a minimum counts of 30 per bin. To test whether binning scheme has a significant impact on our results, we also analyzed the data with two different binning choices: one with a minimum counts of 10 photons per bin, and one with 50. For a minimum counts of 10 (50), we obtained an EW of $22.0\pm5.6$ m\AA\ ($20.4\pm5.7$ m\AA) for the Galactic absorption lines, respectively. These values are consistent with our choice of a minimum 30 counts per bin, suggesting binning scheme does not have a significant impact on our result. We also find the same conclusion for the redshifted line.

We restricted our fits to between 21.3 and 22.6 \AA\ to avoid large scale continuum variation and to ensure an accurate measurement of the continuum near the absorption features that we are interested in. Large scale variations can be caused by various factors such as the uncertainties in the RGS effective area calibration. Overall, we find the continuum level at around 22 \AA\ region varies from 2 to 8\% between narrow band (21.3 and 22.6 \AA)\ and broadband (5 to 30 \AA). Such variations lead to about similar amount of change in the line EW. Again, the errors introduced here are much less than the statistical errors.

\section{Conclusion}

We summarize the major findings of our work here. 

\begin{itemize}

\item We observed the blazar Mkn~501 with the RGS on-board \xmm\ for four times between September 2010 and February 2011, with a total of $\sim 170\ ksec$. The broadband spectra of the four observations can be well-fitted with a power-law with Galactic hydrogen absorption. The photon indices varied between 2 and 2.5, and the continuum flux varied by 30 to 50\% during the four observations. 

\item We detected a redshifted \ion{O}{7} absorption line at around 22.3 \AA\ at 2.5$\sigma$ significance. We estimate the temperature of the absorber is around $10^6$ K. Assuming the absorber is WHIM gas in the foreground Hercules supercluster, the estimated overdensity is $\delta \sim 20$, fully consistent with expectations from cosmological simulations.

\item Alternatively, the redshifted \ion{O}{7} absorber can also be associated with the hot gas in a foreground galaxy along the slightline, or with the blazar itself. Further high resolution X-ray observations and optical surveys of the galaxy distribution will help distinguish between these scenarios. We rule out the possibilities that the absorber is located in the foreground intrgroup medium, intracluster medium, or the halo of the host galaxy.

\item We also clearly detected an \ion{O}{7} absorption line produced by the local ($z=0$) hot gas at high significance ($4\sigma$). The derived line EW and column density are consistent with measurements along other AGN sightlines, suggesting a large covering fraction of metal-enriched hot gas in the Milky Way. Further survey of more AGN sightlines in combining with other observations will help address the location of the hot, X-ray absorbing gas and shed light on the baryon content of the Milky Way.

\end{itemize}

Observing Mkn~501 is an extension of our work on H~2356-309 \citep{buote2009, fang2010}. The primary science goal is to use tracers such as galactic superstructures as a signpost of the WHIM, as suggested by cosmological hydrodynamic simulations, to search for the ``missing baryons". Our success in H~2356-309, the detection in Mkn~501 sightline, as well as \citet{nicastro2013} in which they use UV absorbers as a signpost, suggest this is a much more efficient method in detecting the WHIM. Further observations along this sightline, including both high resolution X-ray observations and optical surveys, will help enhance the statistics of our detection and verify the WHIM nature of the X-ray absorber we detected. 

\acknowledgments
We thank Charles Danforth for the help with the {\sl HST}/GHRS observation of Mkn~501. We also thank the referee for helpful comments and suggestion. TF was partially supported by the National Natural Science Foundation of China under grant No.~11243001 and No.~11273021, also by ``the Fundamental Research Funds for the Central Universities" No.~2013121008.

\end{document}